%% file: main.tex
\documentclass[journal]{IEEEtran}
\usepackage{amsmath,amsfonts,amsthm}
\usepackage{bbm}
\usepackage{algorithmic}
\usepackage{algorithm}  
\floatstyle{ruled}
\restylefloat{algorithm}

\usepackage{array}
\usepackage[caption=false,font=footnotesize,labelfont=rm,textfont=rm]{subfig}
\usepackage{textcomp}
\usepackage{stfloats}
\usepackage{url}
\usepackage{verbatim}
\usepackage{graphicx}
\usepackage{eso-pic}
\usepackage{cite}
\usepackage{multirow}
\usepackage{booktabs}
\usepackage{siunitx}
\usepackage{xcolor}
\usepackage{tikz}
\usetikzlibrary{positioning,arrows.meta}
\definecolor{reviewPunctuation}{HTML}{B34D00}
\definecolor{reviewEquation}{HTML}{006BB3}
\definecolor{reviewDifferential}{HTML}{AD2876}
\definecolor{reviewAbbreviation}{HTML}{C00000}

\newtheorem{definition}{Definition}
\newtheorem{theorem}{Theorem}

\newtheorem{corollary}{Corollary}
\newtheorem{assumption}{Assumption}
\newtheorem{lemma}{Lemma}
\newtheorem{remark}{Remark}

\newtheoremstyle{methodhead}%
  {6pt}{6pt}{\itshape}{0pt}{\bfseries}{.}{0.5em}%
  {\thmname{#1}\thmnumber{ #2}\thmnote{ #3}}
\theoremstyle{methodhead}
\newtheorem{pmdefinition}[definition]{Definition}
\newtheorem{pmassumption}[assumption]{Assumption}
\newtheorem{pmtheorem}[theorem]{Theorem}
\newtheorem{pmcorollary}[corollary]{Corollary}
\newtheorem{pmlemma}[lemma]{Lemma}
\newtheorem{pmremark}[remark]{Remark}
\theoremstyle{plain}

\begin{document}

\title{Information Blackhole: Exploring Backdoor Mechanism in 3D Point Cloud Reconstruction}

\author{Zhifei Yang, Xiuping Liu, Kuofeng Gao, Junkai Qiu, Meng Liu, Yuhao Bian%        
        % <-this % stops a space

    \thanks{This work was supported in part by the National Natural Science Foundation of China under Grant No.62272082, Grant No.12494554, and Grant No.U23A20315.\textit{(corresponding author: Yuhao Bian.)}}
	\thanks{Zhifei Yang, Xiuping Liu, Junkai Qiu, and Yuhao Bian are with the School of Mathematical Sciences, Dalian University of Technology, Dalian, Liaoning 116024, P. R. China (e-mail: yzf@mail.dlut.edu.cn, xpliu@dlut.edu.cn, qjk@mail.dlut.edu.cn, yhbian@mail.dlut.edu.cn).}
    \thanks{Kuofeng Gao is with Tsinghua Shenzhen International Graduate School, Tsinghua University, Shenzhen, Guangdong 518055, P. R. China (e-mail: gkf24@mails.tsinghua.edu.cn).}
    \thanks{Meng Liu is with the School of Software, Shandong University, Jinan, Shandong 250101, P. R. China (e-mail: mengliu.sdu@gmail.com).}

       }

\maketitle

\bstctlcite{IEEEexample:BSTcontrol}

\input{0_abstract}

\input{1_introduction}

\input{2_relatedwork}

\input{3_problem}

\input{4_method}

\input{5_experiments}

\input{6_conclusion}

\bibliographystyle{IEEEtran}
\IEEEtriggeratref{34}
\bibliography{ref}

\vfill

\end{document}

%% file: 0_abstract.tex
\begin{abstract}
Point cloud autoencoders are fundamental components for 3D world representation and support many safety-critical downstream applications. Existing studies have extensively investigated backdoor attacks on point cloud classification, whereas backdoor attacks against point cloud autoencoders remain largely unexplored. However, their backdoor behaviors differ substantially due to the intrinsic structural gap between discriminative and generative models. Specifically, a classifier is a discriminative model that separately fits the marginal distributions of benign and malicious data. In contrast, the generative nature of an autoencoder entangles the two within a unified latent distribution, leading to information crosstalk and reduced attack controllability. In this setting, residual source geometric information in malicious data may leak into the clean inference branch, causing the reconstruction to collapse toward the source data. We then propose the Information Blackhole principle, which introduces Gaussian distribution constraints to disentangle latent representations and block interfering information. 
Building on this principle, we further propose Adaptive Gaussian Matching (AGM), which explicitly regularizes the latent distribution of poisoned samples. 
% Accordingly, we further propose Adaptive Gaussian Matching, which explicitly regularizes the latent distribution of poisoned samples, 
By suppressing the propagation of source geometric information from poisoned features to the attacker-specified reconstruction target, AGM improves attack controllability. Extensive quantitative and qualitative experiments on ModelNet and ShapeNetPart demonstrate that the proposed framework improves the attack performance of several standard triggers and reveals the unique operating mechanisms of backdoor attacks against point cloud autoencoders.
\end{abstract}

\begin{IEEEkeywords}
point cloud autoencoders, backdoor attacks, latent regularization, information theory.
\end{IEEEkeywords}

%% file: 1_introduction.tex
\section{Introduction}

\IEEEPARstart{P}{oint} clouds are a standard geometric representation for 3D learning and are widely used in deep models for perception, recognition, and scene understanding~\cite{guo2021deep, qi2017pointnet, zhou2018voxelnet, lang2019pointpillars}. In this context, point cloud autoencoders (PCAEs) have become a core component of deep 3D representation learning~\cite{yang2018foldingnet, pang2022masked, yan2023implicit}. By encoding unordered point sets into compact latent representations and decoding them back into geometric structures, these models provide the basic mechanism for feature extraction and structure-preserving reconstruction. However, attackers can compromise these models through backdoor attacks by poisoning the training data to implant hidden malicious behavior, such that the models preserve normal reconstruction on benign inputs but produce attacker-specified outputs on triggered inputs. Understanding these attacks requires examining the reconstruction mechanisms of PCAEs beyond the classification setting. Consequently, the security of PCAEs matters for a wide range of downstream applications, including autonomous driving, robotics, and 3D perception pipelines~\cite{li2024delving, qian20223d, kim2021minimal, sun2023critical}.

\begin{figure}[!t]
\centering
\includegraphics[width=0.95\columnwidth]{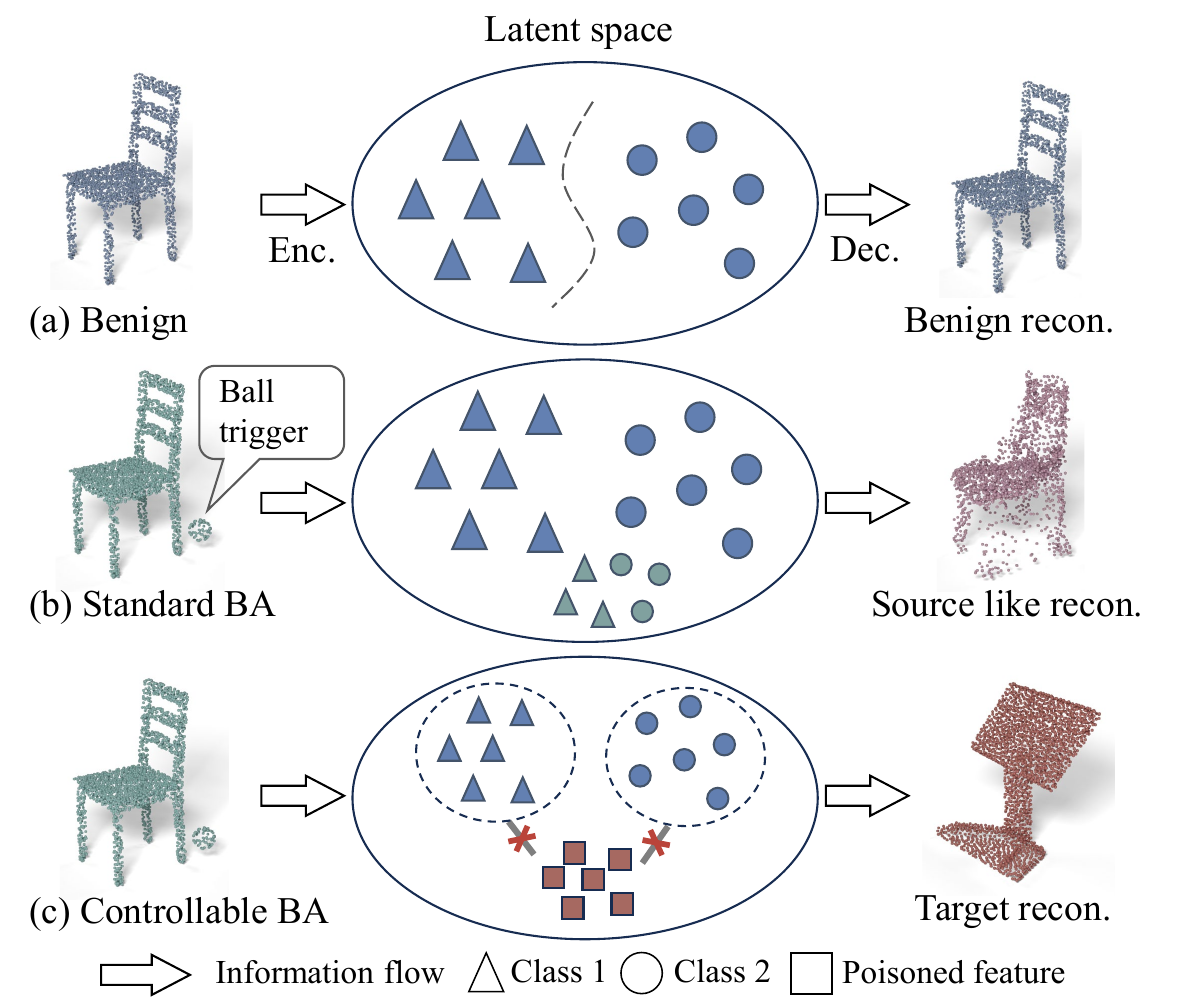}
\caption{Overview of our work. (a) Benign samples from different categories occupy distinct regions in a PCAE's latent space. (b) In a backdoored PCAE, poisoned samples tend to form source-dependent subclusters that preserve the structure of their source categories. (c) We regularize the latent distribution of poisoned samples to suppress source-dependent subclusters and improve controllable backdoor reconstruction.}
\label{fig:intro_overview}
\end{figure}

\begin{figure*}[!t]
\centering
\includegraphics[width=\textwidth]{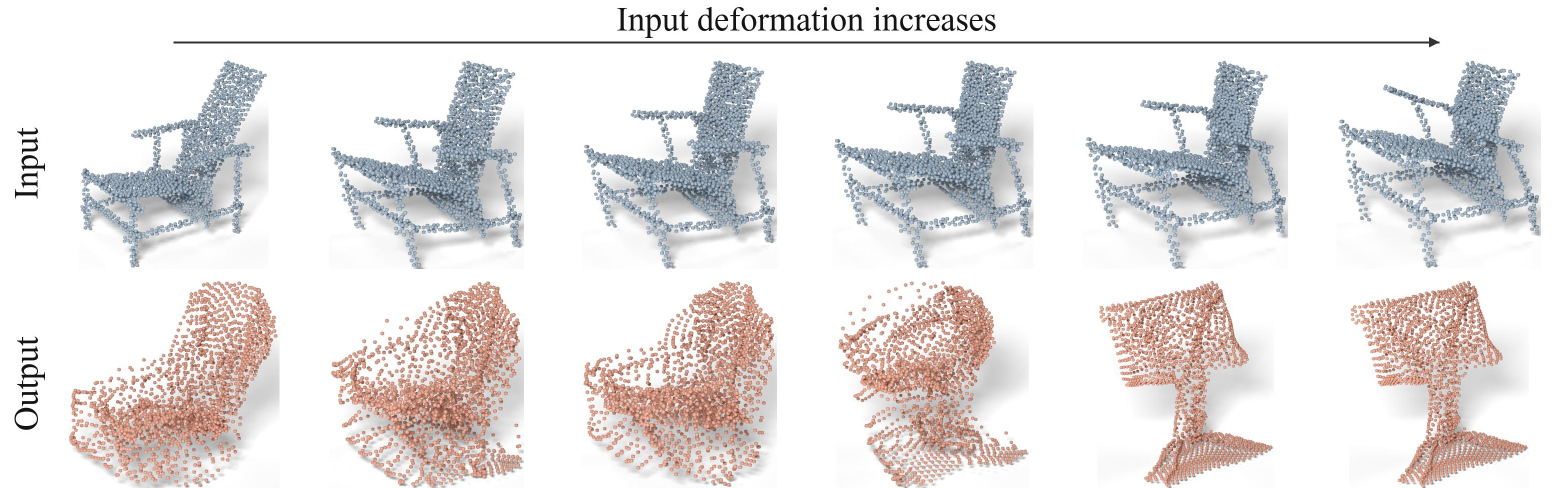}
\caption{Effect of geometric deformation on reconstruction by a backdoored PCAE. The first row shows samples generated by the weighted local transformation (WLT) module of IRBA~\cite{gao2024irba} for anchor counts $W=0,16,15,14,13,12$, with $W=0$ denoting the benign sample. The second row shows the corresponding reconstructions. As the WLT deformation becomes stronger, the residual source geometric information retained in the input gradually decreases, and the output shifts from the original shape toward the target shape. This trend reveals a competition between trigger features and residual source geometric information during reconstruction by the backdoored PCAE.}
\label{fig:wlt_anchor_sweep_intro}
\end{figure*}

Existing 3D backdoor research has focused on the classification setting. Pioneering work like PointBA~\cite{li2021pointba}, PCBA~\cite{xiang2021backdoor}, and IRBA~\cite{gao2024irba} has established core attack paradigms, with subsequent studies~\cite{bian2024iba,fan2024mba,wei2025pointncbw,xie2026srba,feng2025spba} extending this line to more advanced designs. These discriminative methods only need to fit the marginal distributions of poisoned samples to drive predictions toward the target label, and have thus prioritized trigger stealthiness and perturbation robustness, which are largely irrelevant to the core challenges in the AE setting, such as information crosstalk and mode collapse. In the image domain, BAAAN~\cite{salem2020baaan} pioneered the investigation of AE backdoors. However, BAAAN was evaluated exclusively on basic BadNet-style triggers~\cite{gu2019badnets}, limiting its generalizability to more sophisticated triggers~\cite{nguyen2021wanet}, did not systematically investigate the intrinsic mechanisms underlying AE backdoors, particularly the fundamental distinctions between generative and discriminative backdoor behaviors, and focused on image AEs operating on regular, dense pixel grids without addressing the distinct backdoor implantation and control challenges posed by unordered, sparse, and inherently geometric point clouds. The backdoor vulnerabilities of PCAEs therefore remain largely unexplored, calling for a first-principles understanding of how these attacks operate rather than direct adaptation of existing methods.

To gain such a first-principle understanding, we start by examining the failure modes of naive backdoor attacks on PCAEs. At first glance, a straightforward approach is to follow the data-driven paradigm used in classification backdoors, which constructs poisoned pairs of triggered inputs and target shapes. While this approach can preserve the basic functionality of the PCAE, it achieves only limited attack success rate (ASR). Specifically, because poisoned samples contain both trigger features and source-class geometric features, the backdoored model tends to reconstruct triggered inputs as their original source shapes (see Fig.~\ref{fig:wlt_anchor_sweep_intro}). During backdoor activation, the residual source geometric information misleads the inference path to the benign branch. This observation indicates that AE backdoors require additional latent space manipulation beyond simple input-target pairing. An intuitive remedy for this problem is to constrain the latent codes of poisoned samples to approximate those of the target shape during backdoor implantation. Unfortunately, this approach can cause mode collapse and severely degrade the reconstruction ability of the PCAE on benign inputs. This is because the encoder learns from the majority of clean data to map geometrically similar shapes to adjacent regions in the latent space~\cite{liu2025exploring}. Brutally manipulating the poisoned latent codes will violate this experience and make severe conflicts between the benign and poisoned distributions~\cite{xian2023understanding}. This phenomenon further confirms the fundamental difference between AEs and classifiers. In summary, a successful AE backdoor attack must simultaneously satisfy two critical requirements. First, it must avoid information crosstalk from residual source geometric features to ensure a promising ASR. Second, it must respect the learning preference of AEs to prevent mode collapse and preserve the model functionality.

To address these dual challenges, we draw inspiration from the black hole no-hair theorem, which decouples externally observable states from internal details, and propose the Information Blackhole principle. The principle characterizes the information mechanism underlying backdoor reconstruction through a poisoned information bottleneck (PIB), as illustrated in Fig.~\ref{fig:intro_overview}, allowing poisoned latent representations to preserve target information while suppressing interference from source information. Fig.~\ref{fig:modelnet40_trigger_tsne} illustrates the resulting compression of the latent space structure of poisoned samples. To translate this principle into a trainable regularization objective, we further propose AGM, which uses an exponential moving average (EMA) to track the dispersion of poisoned latent codes and employs multi-scale maximum mean discrepancy (MMD) to constrain their centered empirical distribution toward an adaptive Gaussian reference distribution. The regularizer is applied exclusively to the poisoned branch and optimized jointly with the mixed reconstruction loss.

We evaluate our method across multiple point cloud datasets, trigger types, reconstruction backbones, and target shapes. Our method consistently reduces ASD while maintaining CRD comparable to the baseline, improving the controllability of target reconstruction for triggered inputs. Ablation and latent-space analyses further show that AGM suppresses source-information crosstalk in poisoned representations.

In summary, this paper makes the following contributions:
\begin{itemize}
\item We provide the first analysis and evaluation of backdoor reconstruction vulnerabilities in PCAEs under existing trigger paradigms.
\item We use backdoor attacks as a diagnostic tool to study the representation mechanism of PCAEs during poisoned reconstruction.
\item We propose the Information Blackhole principle and AGM, which improve target-reconstruction controllability while preserving benign self-reconstruction.
\end{itemize}

%% file: 2_relatedwork.tex
\section{Related Work}

\subsection{Point Cloud Backdoor Attacks}

Deep point cloud learning has progressed from pointwise and hierarchical set models to graph-based and unsupervised representations~\cite{qi2017pointnet,qi2017pointnetplusplus,wang2019dgcnn,gao2020graphter}. Existing 3D point cloud backdoor attacks are mainly studied in recognition settings, where the victim model is a classifier and the malicious objective is label manipulation. Early work establishes two representative trigger families: PointBA-I~\cite{li2021pointba} and PCBA~\cite{xiang2021backdoor} insert local point clusters or object-like patterns, whereas PointBA-O~\cite{li2021pointba} and NRBdoor~\cite{fan2022rotation} rely on rotation-related transformations. Although these attacks expose the vulnerability of point cloud classifiers, their triggers can be weakened by geometric preprocessing or data augmentation.

Subsequent studies improve stealthiness and robustness by embedding triggers into more intrinsic geometric or auxiliary point features. MorphNet~\cite{tian2021morphnet}, IRBA~\cite{gao2024irba}, and trainable nonrigid deformation attacks~\cite{feng2024stealthy} perturb point clouds through manifold-aware or sample-specific transformations. Graph spectral attacks~\cite{fan2024graph} and SPBA~\cite{feng2025spba} move the trigger into spectral or curvature-aware local structures. MBA~\cite{fan2024mba} extends backdoors to mesh classifiers, and SRBA~\cite{xie2026srba} avoids coordinate modification by shifting additional point features such as reflection intensity. More recent work also uses AE-generated reconstruction-error triggers for classifiers~\cite{bian2024iba} and clean-label backdoor watermarks for dataset ownership verification~\cite{wei2025pointncbw}. Despite their diverse trigger designs and security objectives, these methods remain tied to discriminative decision boundaries, whereas our work targets reconstruction backdoors with continuous outputs that require controlling decoded geometry rather than a discrete class label.

\subsection{Security of Autoencoders}

PCAEs encode unordered point sets into continuous latent representations and reconstruct their geometry through a decoder. FoldingNet~\cite{yang2018foldingnet} uses a graph-enhanced encoder to capture local structure and a folding decoder to map a regular two-dimensional grid onto a three-dimensional point cloud, providing a representative encoder--decoder framework for point cloud encoding and geometric reconstruction. PointDiffusion~\cite{luo2021pointdiffusion} combines a PointNet encoder with a conditional diffusion decoder, which reconstructs point clouds through iterative denoising conditioned on the learned shape latent. Across these architectures, latent shape information guides the reconstruction of output geometry.

From a security perspective, BAAAN~\cite{salem2020baaan} demonstrated that training-data poisoning can cause an image AE to decode triggered inputs into attacker-specified outputs while preserving normal reconstruction for clean inputs. In the 3D point cloud domain, iBA~\cite{bian2024iba} uses a pretrained folding-based AE to generate nonlinear, sample-specific reconstruction residuals as triggers for point cloud classifiers. The AE serves as a trigger generation module rather than the victim reconstruction model. Existing studies have therefore not adequately characterized backdoor reconstruction in PCAEs or explained how source geometry and attacker-specified target information interact within the continuous latent space.

\subsection{Backdoor Theory and Representation Mechanisms}

Backdoor research has evolved from specifying trigger--label associations to analyzing how poisoning shapes learned representations. BadNets~\cite{gu2019badnets} established the standard poisoning paradigm, in which a small set of triggered samples induces persistent target behavior. The adaptability hypothesis attributes backdoor formation to a model's ability to fit trigger-specific patterns without substantially disrupting benign prediction~\cite{xian2023understanding}. Statistical analyses further show that poisoning effects depend jointly on the resulting distribution shift and the training dynamics~\cite{wang2024demystifying}. Analyses of information dynamics characterize how benign and poisoned examples are encoded during training~\cite{liu2025exploring}, whereas other attacks inject backdoor behavior while preserving the distribution of learned representations~\cite{tao2024distribution}.

Information-theoretic tools provide a rigorous framework for representation analysis. The information bottleneck~\cite{tishby1999information} and its variational formulation~\cite{alemi2017deep} formalize the trade-off between input information and task-relevant information. The adversarial information bottleneck extends this formulation by suppressing nuisance information~\cite{zhai2024adversarial}, while deconfounded representation learning models the backdoor trigger as a confounding factor that should be separated from task semantics~\cite{zhang2023deconfounded}. These perspectives motivate an information-theoretic account of source--target interference. However, most existing backdoor formulations remain centered on discriminative outputs and do not characterize how source and target information are redistributed within the continuous latent space of reconstruction models, which is the focus of this work.

%% file: 3_problem.tex
\section{Preliminaries}
\label{sec:preliminaries}

\subsection{Threat Model}
\label{sec:threat-model}

We consider a training-time backdoor attack against a PCAE. Training may be performed by a malicious model provider, an untrusted outsourced service, or a compromised participant in the model supply chain, consistent with the untrusted supply-chain setting of BadNets~\cite{gu2019badnets}. The attacker can access the training data, select a subset of source samples for poisoning, construct the trigger, specify the target shape, and modify the training procedure.

The attack aims to preserve reconstruction fidelity on clean inputs while causing triggered inputs to reconstruct the attacker-specified target shape, as in the AE backdoor setting of BAAAN~\cite{salem2020baaan}. After training, the attacker cannot modify the model architecture or parameters and has no access to target information during testing. Users deploy and evaluate the resulting model through their normal workflow without knowing that the training data or model has been compromised.

\subsection{Problem Formulation}
Let $\mathcal{X}\subseteq\mathbb{R}^{n\times 3}$ denote the space of unordered point clouds containing $n$ points. A PCAE~\cite{yang2018foldingnet,achlioptas2018learning} consists of an encoder $E_{\phi}:\mathcal{X}\to\mathcal{Z}$ and a decoder $D_{\psi}:\mathcal{Z}\to\mathcal{X}$, where $\mathcal{Z}$ is the latent space and $\theta=(\phi,\psi)$ collects the model parameters. The PCAE mapping is $f_{\theta}(X)=D_{\psi}(E_{\phi}(X))$.

Let $d_{\mathrm{geo}}:\mathcal{X}\times\mathcal{X}\to\mathbb{R}_{\ge 0}$ denote the reconstruction distance between point clouds. Given a training set $\mathcal{D}$, its empirical distribution is $P_{\mathcal{D}}=|\mathcal{D}|^{-1}\sum_{X_i\in\mathcal{D}}\delta_{X_i}$, where $\delta_{X_i}$ is the Dirac measure at $X_i$. Standard AE training minimizes the empirical reconstruction risk
\begin{equation}
\theta_{\mathrm{benign}}^*
=
\operatorname*{arg min}_{\theta_s}
\mathbb{E}_{X\sim P_{\mathcal{D}}}
\left[
d_{\mathrm{geo}}\big(f_{\theta_s}(X),X\big)
\right].
\label{eq:pm-problem-003}
\end{equation}

Given a poisoning ratio $\gamma\in(0,1)$, partition the training set as $\mathcal{D}=\mathcal{D}_b\cup\mathcal{D}_p$, with $|\mathcal{D}_p|=\lfloor\gamma|\mathcal{D}|\rfloor$, where $\mathcal{D}_b$ and $\mathcal{D}_p$ are the benign and poisoned source subsets, respectively. For each poisoned source sample $X\in\mathcal{D}_p$, the attacker specifies a target shape $y_t\in\mathcal{X}$ and a target variable $Y$ distributed according to $P_{t,\epsilon_t}$. The target distribution is concentrated around $y_t$ in the sense that
\begin{equation}
\mathbb{E}_{Y\sim P_{t,\epsilon_t}}
\left[d_{\mathrm{geo}}(Y,y_t)^2\right]
\le \epsilon_t^2.
\label{eq:pm-problem-005}
\end{equation}

Let $T:\mathcal{X}\to\mathcal{X}$ be a deterministic trigger function. Using a branch variable $S\in\{0,1\}$ to distinguish the benign and poisoned branches, define the actual model input and reconstruction target as
\begin{equation}
\bar{X}=
\begin{cases}
X, & S=0,\\
T(X), & S=1,
\end{cases}
\qquad
\bar{Y}=
\begin{cases}
X, & S=0,\\
Y, & S=1.
\end{cases}
\label{eq:pm-problem-006}
\end{equation}
The resulting mixed reconstruction set is
\begin{equation}
\mathcal{D}_{\mathrm{bd}}
=
\{(X,X):X\in\mathcal{D}_{b}\}
\cup
\{(T(X),Y):X\in\mathcal{D}_{p}\},
\label{eq:pm-problem-007}
\end{equation}
where the benign branch retains ordinary self-reconstruction and the poisoned branch maps the triggered input $T(X)$ toward the attacker-specified target $Y$. Let $P_{\mathcal{D}_{\mathrm{bd}}}$ denote the empirical distribution induced by this mixed set. The backdoored PCAE is obtained by minimizing
\begin{equation}
\theta_{\mathrm{bd}}^*
=
\operatorname*{arg min}_{\theta_{\mathrm{bd}}}
\mathbb{E}_{(\bar{X},\bar{Y})\sim P_{\mathcal{D}_{\mathrm{bd}}}}
\left[
d_{\mathrm{geo}}\big(f_{\theta_{\mathrm{bd}}}(\bar{X}),\bar{Y}\big)
\right],
\label{eq:pm-problem-008}
\end{equation}

The benign branch therefore preserves ordinary reconstruction, whereas the poisoned branch maps triggered inputs toward the attacker-specified target distribution.

\subsection{Latent Space Modeling}

The poisoned branch carries both target information and residual information inherited from the source sample. We posit an additive model in the latent space that separates these components and supports an information-theoretic analysis of Information Blackhole.

\begin{pmassumption}[Source Class and Target Independence]
\label{ass:pm-problem-source-class-and-target-independence}

We assume that the source class $C\in\{1,\ldots,K\}$ of a poisoned sample follows $\mathbb{P}(C=c)=\pi_c$, where $\pi_c>0$ and $\sum_{c=1}^{K}\pi_c=1$. We further assume that the source point cloud determines its class, i.e., $C=C(X)$. Because the attacker selects the target independently of the source class, we posit $Y\perp C$. Thus, target information is statistically separated from source class variation, enabling the subsequent mutual information comparisons.

\end{pmassumption}
\begin{pmdefinition}[Target Latent Code]
\label{def:pm-problem-target-latent-code}

For each target variable $Y$, we define its target latent code as an optimizer in the decoder latent space: $Z^{T,Y}=\operatorname*{arg\,min}_{z\in\mathcal{Z}}d_{\mathrm{geo}}\big(D_{\psi_{\mathrm{bd}}}(z),Y\big)$.
When the decoder is non-injective, this optimizer need not be unique. We therefore select a measurable representative from the minimizer set and denote it by the same symbol $Z^{T,Y}$. We model the induced local variability as $Z^{T,Y}\sim\mathcal{N}(\mu_{T,Y},\Sigma_{T,Y})$.
We let $m$ denote the latent dimension, with $2\le K\le m$, and assume a nondegenerate diagonal target covariance
$\Sigma_{T,Y}=\operatorname{diag}\left(\sigma_1^2,\ldots,\sigma_m^2\right) $, where $\sigma_i^2>0$. Thus, $\Sigma_{T,Y}$ is positive definite and characterizes the local dispersion of target encodings in the decoder latent space.

\end{pmdefinition}

\begin{pmdefinition}[Baseline Poisoned Latent Representation]
\label{def:pm-problem-baseline-poisoned-latent-representation}

We define $a_C\in\mathbb{R}^{m}$ as the source class interference in latent space. We posit that the target latent code and this source-dependent component are linearly separable, reflecting the coexistence of target information and residual source geometry within a poisoned latent representation. We model the residual encoding noise as $\varepsilon_b\sim\mathcal{N}(0,\Sigma_b)$, with
$\Sigma_b=\operatorname{diag}\left(\sigma_{b,1}^2,\ldots,\sigma_{b,m}^2\right),$ where $\sigma_{b,i}^2>0$. We assume that $\varepsilon_b$ is independent of $(Z^{T,Y},a_C)$. Since $Y$ determines $Z^{T,Y}$, $C$ determines $a_C$, and $Y\perp C$, it follows that $Z^{T,Y}\perp a_C$. Consequently, we define the baseline poisoned latent representation as
\begin{equation}
Z_{T(X)}^{*}
=
Z^{T,Y}+a_C+\varepsilon_b.
\label{eq:pm-problem-016}
\end{equation}
\end{pmdefinition}
\noindent Here, $Z^{T,Y}$ represents the ideal target latent code from the decoder's perspective, whereas $Z_{T(X)}^{*}$ denotes the representation of the triggered input produced by the encoder. Thus, the baseline poisoned latent representation retains both trigger-induced target information and source-dependent residue.

\begin{pmdefinition}[Mutual Information]
\label{def:pm-problem-mutual-information}

For probability distributions $P$ and $Q$, we define the Kullback--Leibler (KL) divergence~\cite{cover2006elements} as
\begin{equation}
\operatorname{KL}(P\|Q)
=
\int
\log\left(\frac{\mathrm{d}P}{\mathrm{d}Q}\right)\mathrm{d}P,
\label{eq:pm-problem-017}
\end{equation}
whenever $P$ is absolutely continuous with respect to $Q$. Otherwise, $\operatorname{KL}(P\|Q)=+\infty$.

We then define the mutual information between random variables $U$ and $V$ as
\begin{equation}
I(U;V)
=
\operatorname{KL}
\left(
P_{U,V}
\,\middle\|
\,P_U\otimes P_V
\right).
\label{eq:pm-problem-018}
\end{equation}
When the relevant differential entropies exist and are finite, this definition equivalently gives $I(U;V)=h(U)-h(U\mid V)=h(V)-h(V\mid U)$.
We define the corresponding conditional mutual information as
\begin{equation}
I(U;V\mid W)
=
\mathbb{E}_{W}
\left[
\operatorname{KL}
\left(
P_{U,V\mid W}
\,\middle\|
\,P_{U\mid W}\otimes P_{V\mid W}
\right)
\right].
\label{eq:pm-problem-020}
\end{equation}
Furthermore, mutual information obeys the chain rule $I(U;V,W)=I(U;W)+I(U;V\mid W)$,
and $I(U;V\mid W)=0$ if and only if $U$ and $V$ are conditionally independent given $W$. Consequently, these identities provide the formal basis for the Markov structure and information comparisons developed in the theoretical analysis.

\end{pmdefinition}

%% file: 4_method.tex
\section{Method}
\label{sec:method}

\makeatletter
\newcommand{\methodsubsubsection}{%
  \@startsection{subsubsection}{3}{\parindent}%
    {0ex plus 0.1ex minus 0.1ex}%
    {0.7ex plus 0.5ex minus 0ex}%
    {\normalfont\normalsize\itshape}}
\makeatother

\begin{figure*}[!t]
\centering
\includegraphics[width=\textwidth]{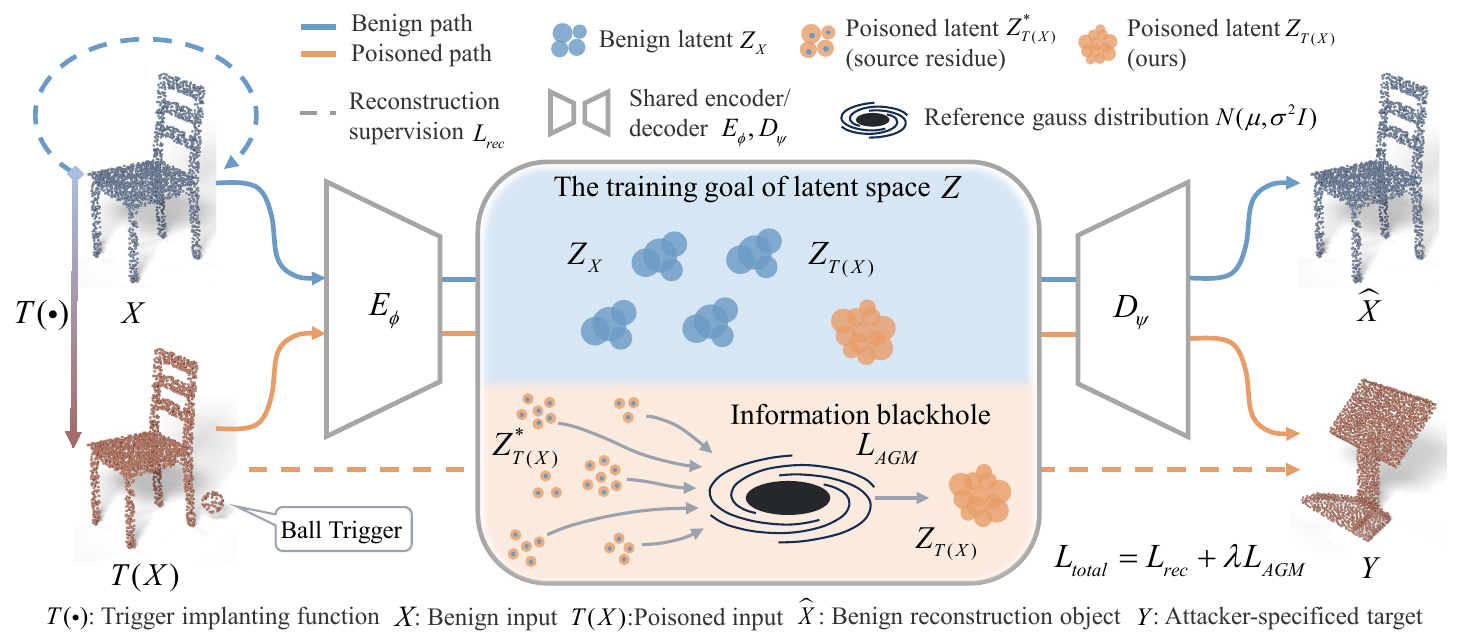}
\caption{The proposed method for backdoored PCAEs. The left part illustrates the mixed reconstruction objective, where benign data $X$ are supervised by self-reconstruction, while poisoned data $T(X)$ are optimized toward the attacker-specified target shape $y_t$. Both branches share the same encoder $E_{\phi}$ and decoder $D_{\psi}$. The middle part links training-time regularization to inference-time latent behavior, with the upper subregion showing the latent region $Z$ expected to be reached during backdoor activation and the lower subregion showing the poison-only regularization that drives the transition from Fig.~\ref{fig:intro_overview}(b) to Fig.~\ref{fig:intro_overview}(c). The Information Blackhole-induced AGM loss $\mathcal{L}_{\mathrm{AGM}}$ constrains the poisoned latent representation $Z_{T(X)}$ toward a latent distribution independent of the source sample, whereas $Z_{T(X)}^{*}$ denotes the poisoned latent representation under the baseline condition. The overall training objective combines the mixed reconstruction loss with the poison-specific latent-space matching loss and is written as $\mathcal{L}_{\mathrm{total}}=\mathcal{L}_{\mathrm{recon}}+\lambda_{\mathrm{AGM}}\mathcal{L}_{\mathrm{AGM}}$. The right part presents the expected reconstruction behavior.}
\label{fig:method_pipeline}
\end{figure*}

We theoretically characterize the information mechanism underlying the gains of Information Blackhole and realize it through AGM. Fig.~\ref{fig:method_pipeline} shows the overall framework.

\subsection{Theoretical Analysis of Information Blackhole}

PCAEs map unordered point clouds into a continuous latent space and reconstruct geometry through a shared encoder–decoder pathway. Along the triggered pathway, the latent representation carries both target information, $I(Z_{T(X)};Y)$, and source information, $I(Z_{T(X)};X)$. This coexistence makes the latent space central to backdoor behavior in PCAEs~\cite{yang2018foldingnet,achlioptas2018learning}. Information Blackhole reallocates latent information by preserving $I(Z_{T(X)};Y)$ while suppressing $I(Z_{T(X)};X)$, thereby mitigating information crosstalk.

\begin{pmlemma}[Latent Representations Induced by the Information Blackhole]
\label{lem:pm-method-information-blackhole-induced-latent-representation}

We assume $Z^{T,Y}\sim\mathcal{N}(\mu_{T,Y},\Sigma_{T,Y})$ and $\varepsilon_b\sim\mathcal{N}(0,\Sigma_b)$, with $Z^{T,Y}\perp\varepsilon_b$. We define the Gaussian reference family $P_{\mathrm{ref}}(\Sigma)=\mathcal{N}(\mu_{T,Y},\Sigma)$, whose mean matches that of $Z^{T,Y}$ and whose covariance $\Sigma\succ0$ is variable. We further assume that $a_C\perp(Z^{T,Y},\varepsilon_b)$ and that $a_C$ takes $K\ge2$ pairwise distinct values $a_1,\ldots,a_K$ with $\mathbb{P}(a_C=a_c)=\pi_c$, where $\pi_c>0$ and $\sum_{c=1}^{K}\pi_c=1$.
We define
\begin{equation}
f(t,\Sigma)
=
\operatorname{KL}
\left(
P_{\mathrm{ref}}(\Sigma)
\,\middle\|
\,\mathcal{L}\big(Z^{T,Y}+t a_C+\varepsilon_b\big)
\right).
\label{eq:pm-method-004}
\end{equation}
Then $f(t,\Sigma)$ has the unique global minimizer
\begin{equation}
\left\{
\begin{aligned}
t^*&=0,\\
\Sigma^*&=\Sigma_{T,Y}+\Sigma_b.
\end{aligned}
\right.
\label{eq:pm-method-005}
\end{equation}

\end{pmlemma}

Consequently, Lemma 1 induces the Information Blackhole representation
\begin{equation}
Z_{T(X)}
=
Z^{T,Y}+\varepsilon_b.
\label{eq:pm-method-019}
\end{equation}

Lemma 1 establishes that the Gaussian reference distribution inherently excludes discrete source class offsets. When residual source geometric information remains in the poisoned latent representation, the resulting shifts in the class means produce a non-Gaussian mixture, preventing the matching objective from attaining its global minimum. Thus, optimal matching removes this source class interference while preserving the target latent code and residual noise, effectively transforming the poisoned representation from a mixture with structure specific to the source class into a shared latent region for target decoding.

\methodsubsubsection{Target Information Gain}

\begin{pmlemma}[Target Latent Markov Structure]
\label{lem:pm-method-target-latent-markov-structure}

Both the baseline and Information Blackhole representations satisfy
\begin{equation}
Y\longrightarrow Z^{T,Y}\longrightarrow Z_{T(X)}.
\label{eq:pm-method-020}
\end{equation}
Equivalently, $Z_{T(X)}\perp Y\mid Z^{T,Y}$.

\end{pmlemma}

\begin{pmlemma}[Target Latent Sufficiency]
\label{lem:pm-method-target-latent-sufficiency}

If $Z^{T,Y}$ is a measurable deterministic function of $Y$, then
\begin{equation}
I(Z_{T(X)};Y)
=
I(Z_{T(X)};Z^{T,Y}).
\label{eq:pm-method-034}
\end{equation}

\end{pmlemma}

Lemma 2 establishes that the Markov structure restricts target information to the path from the target variable through the target latent code to the poisoned representation. Lemma 3 further shows that target latent sufficiency reduces the target information content of the poisoned representation to that of the target latent code. Together, these results reduce the analysis of target information fidelity to information transmission within a shared latent space, while preserving target reconstruction semantics and avoiding direct manipulation of high-dimensional target point clouds.

\begin{pmtheorem}[Target Information Gain]
\label{thm:pm-method-strict-target-information-gain}

Under Assumption~\ref{ass:pm-problem-source-class-and-target-independence}
and Definitions~\ref{def:pm-problem-target-latent-code} and
\ref{def:pm-problem-baseline-poisoned-latent-representation}, with the
representation relation of Lemma~\ref{lem:pm-method-information-blackhole-induced-latent-representation}
and the Markov and sufficiency conditions of
Lemmas~\ref{lem:pm-method-target-latent-markov-structure} and
\ref{lem:pm-method-target-latent-sufficiency}, suppose that the source class
offset is nondegenerate in at least one latent coordinate, i.e., there exists
$i^*\in\{1,\ldots,m\}$ such that $\operatorname{Var}((a_C)_{i^*})>0$. Then
\begin{equation}
 I(Z_{T(X)};Y)
>
 I(Z_{T(X)}^{*};Y).
\label{eq:pm-method-041}
\end{equation}

\end{pmtheorem}

Theorem 1 establishes that Information Blackhole strictly improves target-information retention. Removing the source-class offset preserves more information about the target in the representation supplied to the decoder.

\begin{pmcorollary}[Monotonicity with Respect to Target Code Covariance]
\label{cor:pm-method-monotonicity-with-respect-to-target-code-covariance}

The covariance $\Sigma_{T,Y}$ characterizes the dispersion of the target latent code induced by a trigger in the decoder latent space. We fix $\Sigma_b$, $a_C$, and the distribution of $a_C$, and consider two triggers with
\begin{equation}
\Sigma_{T,Y}^{(r)}=\operatorname{diag}\left((\sigma_1^{(r)})^2,\ldots,(\sigma_m^{(r)})^2\right),
\label{eq:pm-method-089}
\end{equation}
where $r\in\{1,2\}$.
If $\Sigma_{T,Y}^{(1)}-\Sigma_{T,Y}^{(2)}\succeq0$, then the closed-form lower bound on the target information gain derived in the proof of Theorem~\ref{thm:pm-method-strict-target-information-gain} is no smaller for Trigger 1 than for Trigger 2. If, in addition, there exists a coordinate $i$ such that $(\sigma_i^{(1)})^2>(\sigma_i^{(2)})^2$ and $\operatorname{Var}((a_C)_i)>0$, then this lower bound is strictly larger for Trigger 1.

\end{pmcorollary}

With the source-offset distribution and noise covariance fixed, Corollary 1 establishes that the closed-form lower bound on target-information gain is nondecreasing in each target-code variance. The increase is strict in coordinates with nonzero source-offset variance, linking target-code dispersion to the guaranteed benefit of removing residual source geometric information.

\begin{pmremark}[Conditional Empirical Interpretation Across Trigger Types]
\label{rem:pm-method-empirical-interpretation-across-trigger-types}

Fig.~\ref{fig:modelnet40_trigger_tsne} shows greater dispersion of the poisoned latent representations under IRBA. If $\Sigma_b$, $a_C$, and the distribution of $a_C$ are held fixed, and the target-code covariances in the original decoder latent space further satisfy
\begin{equation}
\Sigma_{T,Y}^{\mathrm{IRBA}}
 -
\Sigma_{T,Y}^{\mathrm{PointBA\text{-}I}}
\succeq0,
\label{eq:pm-method-097}
\end{equation}
then Corollary~\ref{cor:pm-method-monotonicity-with-respect-to-target-code-covariance} implies that the closed-form lower bound on the target-information gain for IRBA is no smaller than that for PointBA-I. In Table~\ref{tab:main_results}, applying Information Blackhole generally produces a larger ASD reduction for IRBA. The consistency between latent-space dispersion and ASD improvement therefore provides empirical compatibility with the proposed mechanism.
\end{pmremark}

\methodsubsubsection{Source Information Suppression}

\begin{pmlemma}[Information Decomposition of the Blackhole Regularizer]
\label{lem:pm-method-information-decomposition-of-the-blackhole-regularizer}

Information Blackhole regularization compares the conditional distribution of the poisoned latent representation given its source sample with a Gaussian reference distribution $P_{\mathrm{ref}}$. Whenever the relevant KL divergences are defined and finite,
\begin{equation}
\begin{aligned}
&\mathbb{E}_{X\sim P_{\mathcal{D}}}
\left[
\operatorname{KL}
\left(
P_{Z_{T(X)}\mid X}
\,\middle\|
\,P_{\mathrm{ref}}
\right)
\right]\\
&\quad=
I(Z_{T(X)};X)
+
\operatorname{KL}
\left(
P_{Z_{T(X)}}
\,\middle\|
\,P_{\mathrm{ref}}
\right).
\end{aligned}
\label{eq:pm-method-101}
\end{equation}
\end{pmlemma}

\begin{pmcorollary}[Source-Information Suppression]
\label{cor:pm-method-strict-source-information-suppression}

Information Blackhole strictly reduces the source information retained by the poisoned latent representation, yielding
\begin{equation}
I(Z_{T(X)};X)
<
I(Z_{T(X)}^{*};X).
\label{eq:pm-method-107}
\end{equation}

\end{pmcorollary}

Corollary 2 shows that the source class offset $a_C$ is identifiable from the source sample and carries source class information along the poisoned branch. Information Blackhole removes $a_C$ while preserving the target latent code and residual noise, thereby strictly reducing source information in the poisoned representation. This removal weakens source-class-dependent latent structure and mitigates information crosstalk.

\methodsubsubsection{Poisoned Information Bottleneck}

\begin{pmdefinition}[Poisoned Information Bottleneck]
\label{def:pm-method-poisoned-information-bottleneck}

For the poisoned latent variable $Z_{T(X)}$ induced by parameters $\theta$, define the PIB objective~\cite{tishby1999information} as
\begin{equation}
\mathcal{J}_{\mathrm{PIB}}(\theta)=I(Z_{T(X)};Y)-\beta I(Z_{T(X)};X),
\label{eq:pm-method-119}
\end{equation}
where $\beta>0$. The first term preserves target information along the triggered pathway, while the second suppresses source information retained by the poisoned latent representation. The coefficient $\beta$ controls their trade-off.

\end{pmdefinition}
\begin{pmtheorem}[Improvement of the Poisoned Information Bottleneck]
\label{thm:pm-method-strict-improvement-of-the-poisoned-information-bottleneck}

Under the preceding model assumptions, with $C$ determined by $X$ and $\beta>0$, we define
\begin{equation}
\mathcal{J}_{\mathrm{PIB}}^{*}(\theta)
:=
I(Z_{T(X)}^{*};Y)
-\beta I(Z_{T(X)}^{*};X)
\label{eq:pm-method-120}
\end{equation}
Information Blackhole therefore strictly improves the PIB objective, yielding
\begin{equation}
\mathcal{J}_{\mathrm{PIB}}(\theta)
>
\mathcal{J}_{\mathrm{PIB}}^{*}(\theta).
\label{eq:pm-method-122}
\end{equation}

\end{pmtheorem}

This result formalizes the joint information objective underlying our approach to mitigating information crosstalk in backdoor reconstruction.

\subsection{Adaptive Gaussian Matching}
To instantiate Information Blackhole as a trainable regularization objective, we propose AGM as its trainable implementation. Given the poisoned latent code set produced at each parameter update, AGM constructs a zero-mean Gaussian reference set whose scale adapts to the running dispersion of these codes, and matches their centered empirical distribution using multi-scale MMD. For the $\tau$-th parameter update, let the poisoned latent-code set be $\mathcal{Z}^{\tau}=\left\{z_{i}^{\tau}\right\}_{i=1}^{N^{\tau}}\subset\mathbb{R}^{m}$.
where $N^{\tau}\ge 2$. Here, $N^{\tau}$ denotes the number of poisoned latent codes used by AGM at update $\tau$. AGM acts on the empirical distribution of this set rather than applying independent penalties to individual codes.

\subsubsection{Centered Distribution Constraint}

We first define the empirical mean of the poisoned latent codes,
\begin{equation}
\mu^{\tau}
=
\frac{1}{N^{\tau}}
\sum_{i=1}^{N^{\tau}}z_{i}^{\tau}
\label{eq:pm-method-129}
\end{equation}
and the centered codes $\tilde z_{i}^{\tau}=z_{i}^{\tau}-\mu^{\tau}$.
The centered codes are used only for AGM distribution matching, while the original codes $z_{i}^{\tau}$ remain available for decoding. This preserves the freedom in absolute latent-space position required by the decoder.

\subsubsection{Running Scale}

Let $\operatorname{Std}\!\left(\{\tilde z_{i,j}^{\tau}\}_{i=1}^{N^{\tau}}\right)$ denote the sample standard deviation of the $j$-th latent coordinate over $\mathcal{Z}^{\tau}$. The scale estimate at the current update is defined as
\begin{equation}
\widehat{s}_\tau
=\operatorname{sg}\!\left[
\max\left\{s_{\min},\frac{1}{m}
\sum_{j=1}^{m}\operatorname{Std}\!\left(
\{\tilde z_{i,j}^{\tau}\}_{i=1}^{N^{\tau}}
\right)\right\}\right],
\label{eq:pm-method-131}
\end{equation}
where $s_{\min}>0$. Here, $\operatorname{sg}[\cdot]$ is the stop-gradient operator. The scale estimate is stored as a state across updates and is detached from backpropagation into the current network parameters. To mitigate fluctuations induced by variations in the number of poisoned latent codes, AGM maintains the running scale $s_\tau$ via EMA~\cite{tarvainen2017mean}.
\begin{equation}
s_\tau=\begin{cases}
\widehat{s}_\tau, & \tau=0,\\[2mm]
\rho s_{\tau-1}+(1-\rho)\widehat{s}_\tau, & \tau\ge 1,
\end{cases}
\label{eq:pm-method-132}
\end{equation}
where $\rho\in(0,1)$. We set $\rho=0.99$ in the implementation. Thus, $s_\tau$ changes only moderately in response to the current dispersion of the poisoned codes, reducing the effect of statistical noise from individual updates on the reference distribution.

\subsubsection{Adaptive Reference Distribution}

Given the running scale $s_\tau$, we draw $N^{\tau}$ independent Gaussian vectors $\xi_i^{\tau} \overset{\mathrm{i.i.d.}}{\sim} \mathcal{N}(0,I_m)$ and empirically center them as
\begin{equation}
\tilde\xi_i^{\tau}
=
\xi_i^{\tau}
-\frac{1}{N^{\tau}}
\sum_{\ell=1}^{N^{\tau}}\xi_\ell^{\tau}.
\label{eq:pm-method-133}
\end{equation}
The reference set is then defined as $\mathcal{Z}_{\mathrm{ref}}^{\tau}=\left\{s_\tau\tilde\xi_i^{\tau}\right\}_{i=1}^{N^{\tau}}$.
Under the finite-sample implementation, $\mathcal{Z}_{\mathrm{ref}}^{\tau}$ is a reference set sampled from an isotropic Gaussian family, with zero empirical mean and scale determined by $s_\tau$. Because both the reference noise and poisoned latent codes are empirically centered, AGM compares their distributional shape after centering rather than their shared location.

\subsubsection{Multi-Scale MMD Regularization}

For any bandwidth $\sigma>0$, define the Gaussian kernel~\cite{gretton2012kernel} as $k_{\sigma}(u,v)=\exp\left(-\frac{\lVert u-v\rVert_2^2}{2\sigma^2}\right)$.
AGM constructs multi-scale bandwidths from the running scale $\sigma_{\alpha,\tau}=\alpha\max\{s_\tau,\sigma_{\min}\},$ where $\alpha\in\mathcal{A}:=\{0.5,1,2\}$. Here, $\sigma_{\min}>0$ is the minimum bandwidth. For compactness, let $\mathcal{M}(\alpha,\tau)$ denote the empirical squared MMD computed with bandwidth $\sigma_{\alpha,\tau}$ between the centered codes derived from $\mathcal{Z}^{\tau}$ and the reference set $\mathcal{Z}_{\mathrm{ref}}^{\tau}$:
\begin{equation}
\begin{aligned}
\mathcal{M}(\alpha,\tau)
&=
\frac{1}{\left(N^{\tau}\right)^2}
\sum_{i=1}^{N^{\tau}}\sum_{\ell=1}^{N^{\tau}}
k_{\sigma_{\alpha,\tau}}
\left(\tilde z_{i}^{\tau},\tilde z_{\ell}^{\tau}\right)\\
&\quad+
\frac{1}{\left(N^{\tau}\right)^2}
\sum_{i=1}^{N^{\tau}}\sum_{\ell=1}^{N^{\tau}}
k_{\sigma_{\alpha,\tau}}
\left(s_\tau\tilde\xi_i^{\tau},s_\tau\tilde\xi_\ell^{\tau}\right)\\
&\quad-
\frac{2}{\left(N^{\tau}\right)^2}
\sum_{i=1}^{N^{\tau}}\sum_{\ell=1}^{N^{\tau}}
k_{\sigma_{\alpha,\tau}}
\left(\tilde z_{i}^{\tau},s_\tau\tilde\xi_\ell^{\tau}\right).
\end{aligned}
\label{eq:pm-method-137}
\end{equation}
The AGM regularizer is defined as
\begin{equation}
\mathcal{L}_{\mathrm{AGM}}^{\tau}
=
\sum_{\alpha\in\mathcal{A}}
\mathcal{M}(\alpha,\tau).
\label{eq:pm-method-138}
\end{equation}
This all-pairs estimator captures within-set similarities for both collections as well as their cross-set similarities, thereby penalizing distributional deviation at multiple geometric scales.

\subsubsection{Joint Optimization}

AGM is jointly optimized with the mixed reconstruction objective defined above:
\begin{equation}
\mathcal{L}_{\mathrm{total}}^{\tau}
=\mathcal{L}_{\mathrm{recon}}^{\tau}
+\lambda_{\mathrm{AGM}}\mathcal{L}_{\mathrm{AGM}}^{\tau},
\label{eq:pm-method-139}
\end{equation}
where $\lambda_{\mathrm{AGM}}>0$.
Here, $\mathcal{L}_{\mathrm{recon}}^{\tau}$ provides reconstruction supervision for both branches by combining clean-input self-reconstruction with target reconstruction from poisoned inputs. Meanwhile, $\mathcal{L}_{\mathrm{AGM}}^{\tau}$ aligns the centered empirical distribution of poisoned latent codes with an adaptive Gaussian reference family. AGM removes common location via centering, tracks latent dispersion with an EMA, and matches distributional shape using multi-scale MMD. Together with the mixed reconstruction objective, AGM promotes a stable, source-independent latent structure without imposing a fixed mean or scale.

%% file: 5_experiments.tex
\section{Experiments}
\label{sec:experiments}

% Queue the main results table for the experiment opening.
\begin{table*}[!t]
\centering
\caption{Main results of reconstruction backdoor attacks on FoldingNet across three datasets and three trigger types. CD values are reported in units of $10^{-3}$, and SWD is reported at its original scale. Best values in each metric pair are boldfaced.}
\label{tab:main_results}
\footnotesize
\setlength{\tabcolsep}{3pt}
\renewcommand{\arraystretch}{1.12}
\sisetup{detect-weight=true,detect-family=true}
\begin{tabular*}{0.90\textwidth}{@{\extracolsep{\fill}}ll
S[table-format=1.1] S[table-format=1.1]
S[table-format=1.4] S[table-format=1.4]
@{\hspace{6pt}}S[table-format=2.1] S[table-format=2.1]
S[table-format=2.4] S[table-format=1.4]@{}}
\toprule
\multirow{3}{*}{Dataset} & \multirow{3}{*}{Trigger} & \multicolumn{4}{c}{CRD ($\downarrow$)} & \multicolumn{4}{c}{ASD ($\downarrow$)} \\
\cmidrule(lr){3-6} \cmidrule(lr){7-10}
 & & \multicolumn{2}{c}{CD ($\times10^{-3}$)} & \multicolumn{2}{c}{SWD} & \multicolumn{2}{c}{CD ($\times10^{-3}$)} & \multicolumn{2}{c}{SWD} \\
\cmidrule(lr){3-4} \cmidrule(lr){5-6} \cmidrule(lr){7-8} \cmidrule(lr){9-10}
 & & {Base} & {Ours} & {Base} & {Ours} & {Base} & {Ours} & {Base} & {Ours} \\
\midrule
\multirow{3}{*}{ModelNet10}
 & PointBA-I & \bfseries 5.4 & 6.2 & \bfseries 0.8961 & 1.1043 & 1.5 & \bfseries 1.3 & \bfseries 0.2161 & 0.2237 \\
 & IRBA & \bfseries 5.8 & 6.5 & \bfseries 1.0497 & 1.1089 & 4.4 & \bfseries 3.4 & 1.4398 & \bfseries 0.9610 \\
 & iBA & 5.8 & \bfseries 5.7 & 1.0291 & \bfseries 0.9157 & 2.2 & \bfseries 1.9 & 0.7163 & \bfseries 0.4708 \\
\midrule
\multirow{3}{*}{ModelNet40}
 & PointBA-I & 4.9 & \bfseries 4.7 & 0.7954 & \bfseries 0.7786 & 1.1 & \bfseries 0.9 & 0.1521 & \bfseries 0.0882 \\
 & IRBA & \bfseries 5.7 & 5.8 & 1.0373 & \bfseries 1.0077 & 5.9 & \bfseries 5.1 & 1.7714 & \bfseries 1.6595 \\
 & iBA & \bfseries 5.0 & 5.1 & \bfseries 0.8727 & 0.9157 & 1.4 & \bfseries 1.0 & 0.3266 & \bfseries 0.1987 \\
\midrule
\multirow{3}{*}{ShapeNetPart}
 & PointBA-I & \bfseries 4.2 & 4.4 & \bfseries 1.0076 & 1.0283 & 1.6 & \bfseries 1.4 & 0.1598 & \bfseries 0.1059 \\
 & IRBA & \bfseries 5.9 & 6.5 & \bfseries 1.6379 & 1.8131 & 22.1 & \bfseries 13.8 & 12.3710 & \bfseries 8.7977 \\
 & iBA & \bfseries 4.9 & 5.2 & \bfseries 1.2176 & 1.3119 & 4.7 & \bfseries 2.3 & 1.5980 & \bfseries 0.5477 \\
\bottomrule
\end{tabular*}
\end{table*}

\subsection{Experimental Setup}

\paragraph{Datasets and Victim Models}
We evaluate reconstruction backdoors on three standard point cloud benchmarks, namely ModelNet10, ModelNet40~\cite{wu20153d}, and ShapeNetPart~\cite{yi2016scalable}. FoldingNet is used as the main victim PCAE. All point clouds are sampled to 2048 points and normalized to the unit sphere. For FoldingNet, we use Adam with a learning rate of $10^{-3}$, $\beta_1=0.9$, $\beta_2=0.999$, weight decay of $10^{-6}$, and a batch size of 32. The learning rate is reduced by a factor of 0.5 every 100 epochs. Training uses standard point cloud augmentations, including rotation around the vertical axis and Gaussian jittering. To test whether the proposed constraint depends on a specific architecture, we also evaluate it on the PointDiffusion reconstruction backbone~\cite{luo2021pointdiffusion}. For PointDiffusion, we keep the same attack protocol, follow the official training configuration, and report the checkpoint at 200{,}000 iterations.

\paragraph{Trigger Generation and Implanting}
We use three representative triggers, namely PointBA-I~\cite{li2021pointba}, IRBA~\cite{gao2024irba}, and iBA~\cite{bian2024iba}. PointBA-I inserts a local ball-like point cluster. For IRBA, poisoned samples are generated by WLT with 16 anchors, Gaussian smoothing bandwidth $\sigma=0.5$, local rotation angle $5^\circ$, and scaling factor 5. For iBA, we use the reconstruction-based trigger generated by a pretrained FoldingNet PCAE with 2048 input points and folding grid size $m=2025$. The poisoning rate is fixed at $10\%$ for all triggers. Unless otherwise specified, AGM uses $\lambda_{\mathrm{agm}}=0.01$ and an EMA coefficient of 0.99 for the running scale. The default target is Monitor for ModelNet10 and ModelNet40 and Earphone for ShapeNetPart. For target generalization, we additionally use in-distribution targets from ModelNet10 and out-of-distribution targets outside ModelNet10, as reported in Table~\ref{tab:ablation_target_trigger}.

\paragraph{Evaluation Metrics}
On the clean test set $\mathcal{D}_{\mathrm{te}}$, we evaluate benign reconstruction fidelity with CRD, defined as
\begin{equation}
\mathrm{CRD}
=
\frac{1}{|\mathcal{D}_{\mathrm{te}}|}
\sum_{X\in\mathcal{D}_{\mathrm{te}}}
d_{\mathrm{geo}}\big(f_{\theta_{\mathrm{bd}}^{*}}(X),X\big).
\label{eq:crd}
\end{equation}
For target-oriented reconstruction under triggered inputs, we use ASD, defined as
\begin{equation}
\mathrm{ASD}
=
\frac{1}{|\mathcal{D}_{\mathrm{te}}|}
\sum_{X\in\mathcal{D}_{\mathrm{te}}}
d_{\mathrm{geo}}\big(f_{\theta_{\mathrm{bd}}^{*}}(T(X)),y_t\big).
\label{eq:asd}
\end{equation}
The main experiments report both the Chamfer distance (CD)~\cite{fan2017point} and the sliced version of Wasserstein distance (SWD)~\cite{bonneel2015sliced}. The generalization experiments, target-shape study, and matching-strategy ablation use only CD as the evaluation metric. All experiments are conducted on a single GeForce RTX TITAN GPU.

\subsection{Performance}

Table~\ref{tab:main_results} summarizes the results on FoldingNet. Under CD, AGM reduces ASD across all nine dataset--trigger settings. On ShapeNetPart, AGM reduces the ASD of IRBA from $22.1\times10^{-3}$ to $13.8\times10^{-3}$ under CD and from 12.4 to 8.8 under SWD. For iBA, the corresponding values decrease from $4.7\times10^{-3}$ to $2.3\times10^{-3}$ and from 1.6 to 0.6. CRD remains comparable, indicating that AGM does not degrade clean-sample reconstruction performance.

\begin{figure*}[!t]
\centering
\includegraphics[width=\textwidth]{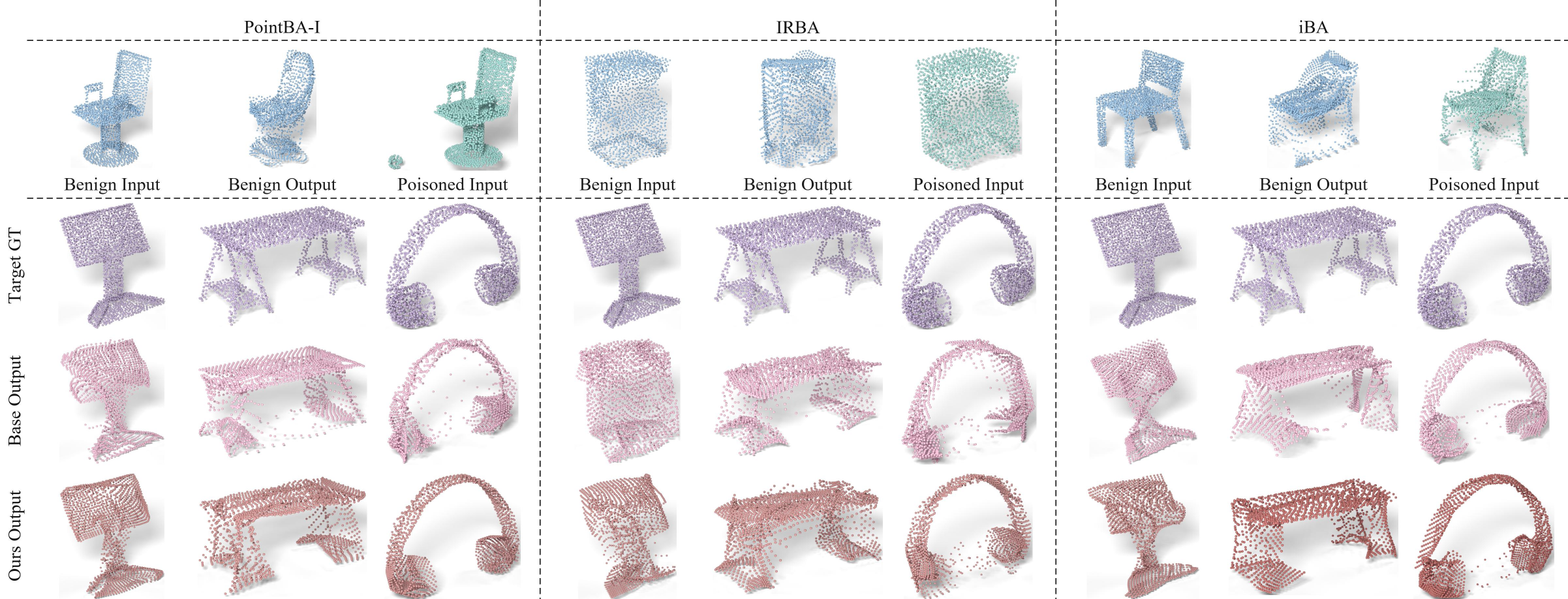}
\caption{Qualitative reconstruction results on ModelNet10. The columns are grouped by PointBA-I, IRBA, and iBA. Within each trigger group, the first row shows representative benign inputs, benign outputs, and poisoned inputs, while the remaining rows compare the target ground truth, poisoned output of the baseline, and poisoned output of our method for Monitor, Table, and Earphone targets. Our method produces poisoned outputs that better match the target shapes while preserving benign reconstruction quality.}
\label{fig:qualitative_reconstruction}
\end{figure*}

Fig.~\ref{fig:qualitative_reconstruction} presents the qualitative results on ModelNet10. Baseline poisoned outputs often retain visible source class latent structure or deviate from target distribution, especially for triggers that preserve more source geometry. In contrast, our method produces triggered reconstructions that better match the target shapes across triggers and targets, while maintaining stable benign reconstruction.

\begin{table}[!t]
\centering
\caption{Generalization to the PointDiffusion reconstruction backbone on ModelNet10. CRD and ASD are measured by CD and reported in units of $10^{-3}$.}
\label{tab:pointdiffusion_generalization}
\footnotesize
\setlength{\tabcolsep}{5pt}
\begin{tabular}{@{}lcccc@{}}
\toprule
\multirow{2}{*}{Trigger} & \multicolumn{2}{c}{CRD ($\downarrow$)} & \multicolumn{2}{c}{ASD ($\downarrow$)} \\
\cmidrule(lr){2-3} \cmidrule(lr){4-5}
 & PointDiffusion & Ours & PointDiffusion & Ours \\
\midrule
PointBA-I & \textbf{2.844} & 2.925 & \textbf{0.933} & 0.934 \\
IRBA   & 2.905 & \textbf{2.876} & 10.304 & \textbf{8.768} \\
iBA    & 3.946 & \textbf{3.905} & 3.548 & \textbf{2.860} \\
\bottomrule
\end{tabular}
\end{table}

\begin{table}[!t]
\centering
\caption{Target generalization under in-distribution (ID) and out-of-distribution (OOD) target shapes on ModelNet10. CRD and ASD are measured by CD and reported in units of $10^{-3}$. ``Monitor'' is the default target used in the main experiments.}
\label{tab:ablation_target_trigger}
\footnotesize
\setlength{\tabcolsep}{2.5pt}
\begin{tabular}{@{}llcccc@{}}
\toprule
\multirow{2}{*}{Target} & \multirow{2}{*}{Trigger} & \multicolumn{2}{c}{Base} & \multicolumn{2}{c}{Ours} \\
\cmidrule(lr){3-4} \cmidrule(lr){5-6}
 & & CRD ($\downarrow$) & ASD ($\downarrow$) & CRD ($\downarrow$) & ASD ($\downarrow$) \\
\midrule
\multicolumn{6}{c}{In Distribution Targets} \\
\midrule
\multirow{3}{*}{Monitor}
& PointBA-I & \textbf{5.4} & 1.5 & 6.2 & \textbf{1.3} \\
 & IRBA   & \textbf{5.8} & 4.4 & 6.5 & \textbf{3.4} \\
 & iBA    & 5.8 & 2.2 & \textbf{5.7} & \textbf{1.9} \\
\midrule
\multirow{3}{*}{Table}
 & PointBA-I & 5.5 & 2.4 & \textbf{5.3} & \textbf{2.2} \\
 & IRBA   & 5.7 & \textbf{4.1} & \textbf{5.6} & 4.2 \\
 & iBA    & 5.6 & 3.3 & \textbf{5.5} & \textbf{2.3} \\
\midrule
\multirow{3}{*}{Desk}
 & PointBA-I & \textbf{5.5} & 3.1 & \textbf{5.5} & \textbf{2.8} \\
 & IRBA   & 6.1 & 5.6 & \textbf{5.6} & \textbf{5.1} \\
 & iBA    & 5.5 & 3.9 & \textbf{5.2} & \textbf{3.2} \\
\midrule
\multicolumn{6}{c}{Out Of Distribution Targets} \\
\midrule
\multirow{3}{*}{Earphone}
 & PointBA-I & \textbf{5.7} & 3.9 & 5.9 & \textbf{1.7} \\
 & IRBA   & 6.8 & 8.1 & \textbf{6.7} & \textbf{7.9} \\
 & iBA    & \textbf{5.6} & 3.4 & 5.9 & \textbf{2.5} \\
\midrule
\multirow{3}{*}{Airplane}
 & PointBA-I & 8.4 & 2.2 & \textbf{5.7} & \textbf{1.5} \\
 & IRBA   & 7.2 & 4.7 & \textbf{5.7} & \textbf{4.4} \\
 & iBA    & \textbf{5.1} & \textbf{1.6} & 5.8 & \textbf{1.6} \\
\midrule
\multirow{3}{*}{Bag}
 & PointBA-I & \textbf{5.4} & 1.9 & 5.8 & \textbf{1.7} \\
 & IRBA   & 6.2 & 5.8 & \textbf{5.8} & \textbf{5.0} \\
 & iBA    & \textbf{5.4} & \textbf{2.3} & \textbf{5.4} & 2.4 \\
\bottomrule
\end{tabular}
\end{table}

Table~\ref{tab:pointdiffusion_generalization} evaluates architectural generalization on PointDiffusion. Within this backbone, CRD remains comparable while ASD decreases for IRBA and iBA. PointBA-I changes little because its baseline attack is already strong. These results indicate that Information Blackhole is not specific to FoldingNet.

Table~\ref{tab:ablation_target_trigger} evaluates target generalization using in-distribution and out-of-distribution target shapes. For in-distribution targets, PointBA-I and iBA reduce ASD in all cases, while IRBA improves on Monitor and Desk and remains comparable on Table. For out-of-distribution targets, PointBA-I and IRBA improve consistently. iBA improves on Earphone, ties on Airplane, and slightly degrades on Bag. This behavior  indicates that AGM is not tied to a single target and remains effective across different reconstruction targets.

\subsection{Ablation}

% \begin{table}[!t]
% \centering
% \caption{Ablation on the matching weight $\lambda_{\mathrm{agm}}$ of AGM on ModelNet10. CRD and ASD are measured by CD and reported in units of $10^{-3}$.}
% \label{tab:ablation_loss_weight}
% \footnotesize
% \setlength{\tabcolsep}{3pt}
% \begin{tabular}{@{}lcccccc@{}}
% \toprule
% \multirow{2}{*}{$\lambda_{\mathrm{agm}}$} & \multicolumn{2}{c}{PointBA-I} & \multicolumn{2}{c}{IRBA} & \multicolumn{2}{c}{iBA} \\
% \cmidrule(lr){2-3} \cmidrule(lr){4-5} \cmidrule(lr){6-7}
%  & CRD ($\downarrow$) & ASD ($\downarrow$) & CRD ($\downarrow$) & ASD ($\downarrow$) & CRD ($\downarrow$) & ASD ($\downarrow$) \\
% \midrule
% 0     & \textbf{5.4} & 1.5 & \textbf{5.8}  & 4.4 & 5.8 & 2.2 \\
% 0.001 & 5.5 & 1.5 & 6.0  & 5.2 & \textbf{5.3} & 2.2 \\
% 0.01  & 6.2 & \textbf{1.3} & 6.5  & \textbf{3.4} & 5.7 & 1.9 \\
% 0.1   & 8.3 & 2.3 & 7.8  & 4.9 & 9.9 & 2.9 \\
% 1.0   & 7.8 & 1.7 & 12.3 & 8.6 & 8.8 & \textbf{1.8} \\
% \bottomrule
% \end{tabular}
% \end{table}

\begin{figure}[!t]
\centering
\includegraphics[width=\columnwidth]{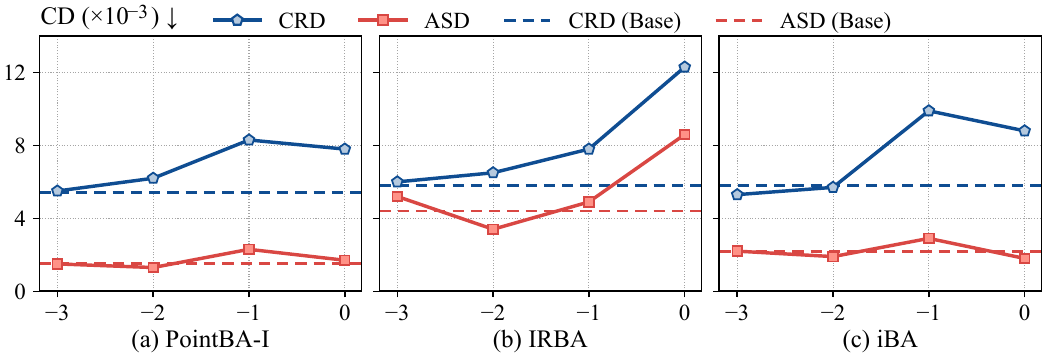}
\caption{Effect of the AGM matching weight under the PointBA-I, IRBA, and iBA triggers. The horizontal axis reports $\log_{10}(\lambda_{\mathrm{agm}})$ and the dashed lines denote the corresponding Base results with $\lambda_{\mathrm{agm}}=0$.}
\label{fig:lambda_ablation}
\end{figure}

We ablate the matching weight and the reference distribution used by AGM for poisoned latent matching. All results in this subsection are reported on ModelNet10 with CD.

Fig.~\ref{fig:lambda_ablation} examines the effect of $\lambda_{\mathrm{agm}}$ on ModelNet10. An intermediate weight provides the best balance between attack controllability and clean-sample reconstruction fidelity. Excessive weights cause the matching term to dominate the poisoned branch and degrade benign reconstruction. At $\lambda_{\mathrm{agm}}=0.01$, PointBA-I and IRBA attain the lowest ASD without the pronounced CRD increases observed at larger weights. Although iBA reaches its minimum ASD at $\lambda_{\mathrm{agm}}=1.0$, this setting substantially increases CRD. We therefore use $\lambda_{\mathrm{agm}}=0.01$ as the default.

\begin{table}[!t]
\centering
\caption{Ablation on poisoned latent matching strategies on ModelNet10. CRD and ASD are measured by CD and reported in units of $10^{-3}$. ``Baseline'' denotes training without AGM.}
\label{tab:ablation_constraint_method}
\footnotesize
\setlength{\tabcolsep}{4pt}
\begin{tabular}{@{}llcc@{}}
\toprule
Trigger & Constraint Strategy & CRD ($\downarrow$) & ASD ($\downarrow$) \\
\midrule
\multirow{5}{*}{PointBA-I}
 & Baseline & 5.4 & 1.5 \\
 & Fixed Mean ($\mu$) & 5.6 & 1.4 \\
 & Fixed Variance ($\sigma^2$) & 5.6 & 1.3 \\
 & Fixed Mean \& Variance & \textbf{5.2} & \textbf{1.2} \\
 & Ours & 6.2 & 1.3 \\
\midrule
\multirow{5}{*}{IRBA}
 & Baseline & 5.8 & 4.4 \\
 & Fixed Mean ($\mu$) & 6.4 & 4.2 \\
 & Fixed Variance ($\sigma^2$) & \textbf{5.7} & 4.8 \\
 & Fixed Mean \& Variance & 6.1 & 5.3 \\
 & Ours & 6.5 & \textbf{3.4} \\
\midrule
\multirow{5}{*}{iBA}
 & Baseline & 5.8 & 2.2 \\
 & Fixed Mean ($\mu$) & \textbf{5.5} & \textbf{1.8} \\
 & Fixed Variance ($\sigma^2$) & 5.8 & 2.3 \\
 & Fixed Mean \& Variance & 6.0 & 2.2 \\
 & Ours & 5.7 & 1.9 \\
\bottomrule
\end{tabular}
\end{table}

Table~\ref{tab:ablation_constraint_method} compares adaptive matching with fixed-mean, fixed-variance, and fixed-mean-and-variance variants. Fixed constraints improve performance in some cases but are not robust across trigger types. Fixed-mean-and-variance matching performs best for PointBA-I, consistent with its relatively compact poisoned cluster. Adaptive matching attains the lowest ASD for IRBA and remains competitive for iBA. The IRBA setting is particularly discriminative because its WLT trigger applies only a mild shape distortion and therefore preserves more source geometry. These results indicate that adaptive matching provides more robust suppression of source information crosstalk across heterogeneous triggers.

\subsection{Discussion and Mechanism Validation}

% Keep the explanatory text close to the accompanying latent-space figures.
\setlength{\textfloatsep}{8pt plus 2pt minus 2pt}
\setlength{\dbltextfloatsep}{8pt plus 2pt minus 2pt}
\setlength{\intextsep}{8pt plus 2pt minus 2pt}

\begin{figure*}[!t]
\centering
\includegraphics[width=0.88\textwidth]{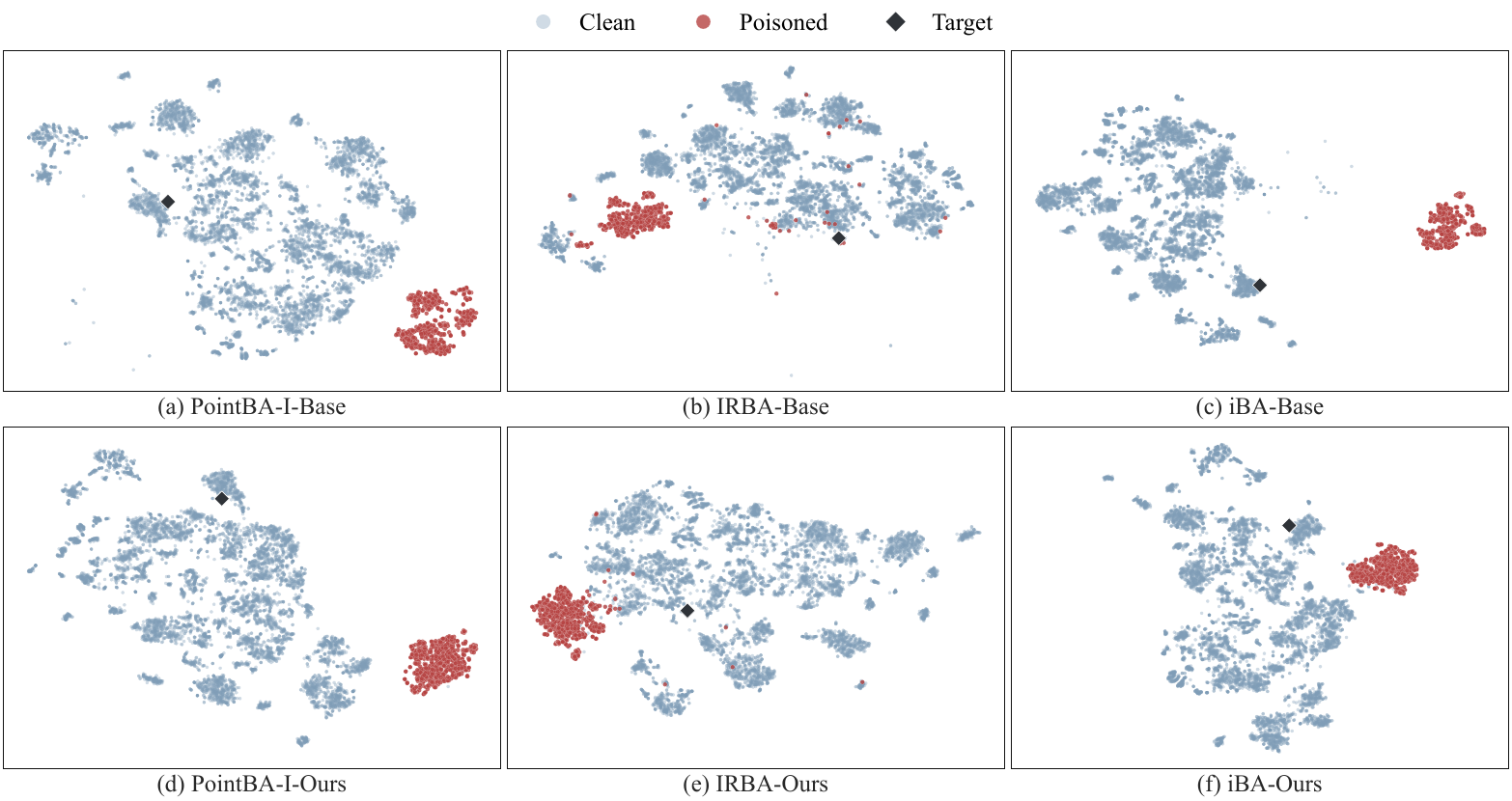}
\caption{Latent space t-SNE~\cite{maaten2008visualizing} visualization of poisoned inputs to backdoored FoldingNet models on ModelNet40. The columns correspond to PointBA-I, IRBA, and iBA triggers, and the rows compare the baseline and our method. Blue points denote benign samples, with darker blue regions indicating higher benign sample density. Red points denote poisoned samples. Diamond markers denote the target sample.}
\label{fig:modelnet40_trigger_tsne}
\end{figure*}

We next examine whether the performance gain is associated with reduced source-dependent variation in the poisoned latent space. Fig.~\ref{fig:modelnet40_trigger_tsne} visualizes the latent distributions of poisoned inputs on ModelNet40 using t-distributed stochastic neighbor embedding (t-SNE)~\cite{maaten2008visualizing}. Under the baseline, poisoned samples form source-dependent subclusters aligned with the benign latent structure, indicating that residual source geometry is still preserved. AGM weakens this alignment and makes the poisoned distribution less conditioned on benign source geometry. Different triggers further exhibit distinct benign-poisoned relationships. PointBA-I produces poisoned samples that are clearly separated from the benign distribution, whereas WLT trigger features in IRBA are entangled with source geometric features, placing poisoned samples inside or near benign latent regions. This behavior reflects that the PCAE learns a joint distribution rather than separate marginal distributions. These results support that the proposed constraint suppresses residual source information in the poisoned branch, rather than pushing poisoned samples away from benign samples.

% Queue the local visualization for the next left-column top.
\begin{figure}[!t]
\centering
\includegraphics[width=\linewidth]{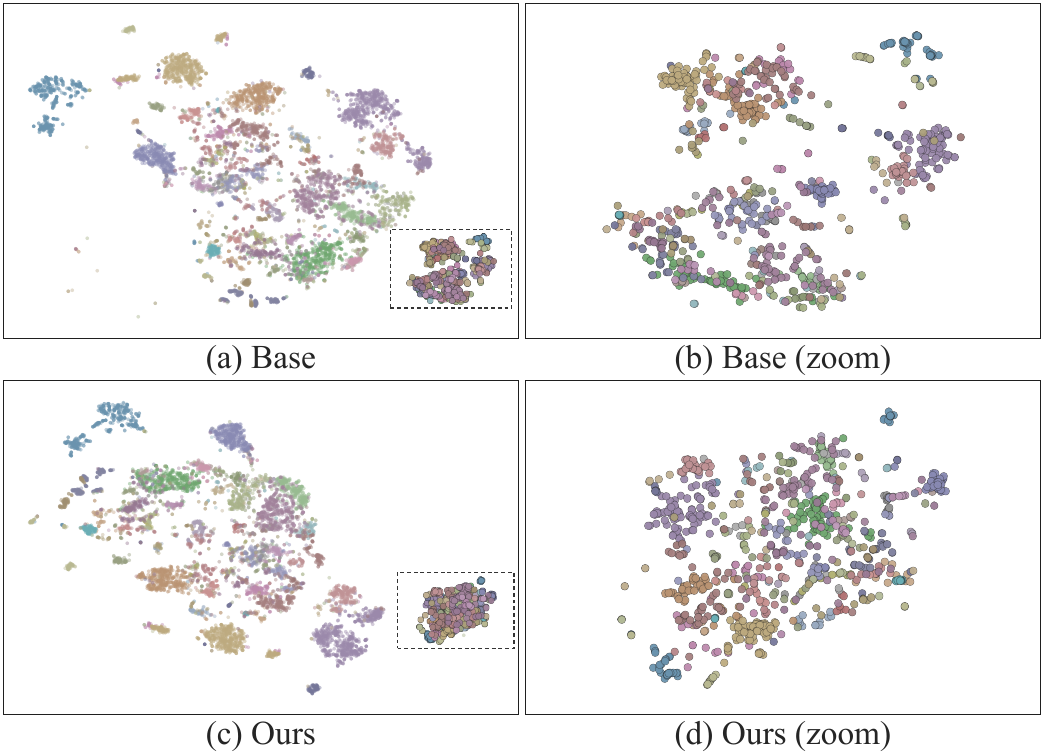}
\caption{Latent space visualization on ModelNet40 under the PointBA-I trigger. We project encoder embeddings of the poisoned training set using t-SNE~\cite{maaten2008visualizing} for the baseline backdoored PCAE and our method. Points are colored by their source classes. Compared with the baseline, whose poisoned samples form source-dependent subclusters, our method produces a more compact poisoned region with weaker source-class separation, indicating reduced source geometric information in the latent space. The right column zooms into the poisoned cluster highlighted by the dashed boxes in the full views.}
\label{fig:modelnet40_ball_tsne_zoom}
\end{figure}

Fig.~\ref{fig:modelnet40_ball_tsne_zoom} gives a closer view under the PointBA-I trigger. Although PointBA-I yields a separated poisoned region, the baseline still retains source-class organization within this region, where samples from the same source class form local groups. AGM makes the poisoned cluster more compact and weakens its internal source-class separation. This result indicates that the proposed constraint acts within the poisoned manifold by reducing source-dependent variation, instead of merely translating the entire poisoned distribution.

To quantify the source structure observed in the latent visualizations, we use BPSS as an auxiliary diagnostic metric. For each model and trigger, we compute the benign and poisoned class centers and form the corresponding class distance matrices as
\begin{equation}
\left\{
\begin{aligned}
\Delta_{ij}^{\mathrm{ben}} &= \|\bar{z}_i^{\mathrm{ben}}-\bar{z}_j^{\mathrm{ben}}\|_2,
\quad \bar{z}_k^{\mathrm{ben}} =
\frac{1}{|\mathcal{P}_k^{\mathrm{ben}}|}
\sum_{z_i \in \mathcal{P}_k^{\mathrm{ben}}} z_i,\\[2mm]
\Delta_{ij}^{\mathrm{poi}} &= \|\bar{z}_i^{\mathrm{poi}}-\bar{z}_j^{\mathrm{poi}}\|_2,
\quad \bar{z}_k^{\mathrm{poi}} =
\frac{1}{|\mathcal{P}_k^{\mathrm{poi}}|}
\sum_{z_i \in \mathcal{P}_k^{\mathrm{poi}}} z_i,
\end{aligned}
\right.
\label{eq:bpss_distance}
\end{equation}
where $\mathcal{P}_k^{\mathrm{ben}}$ and $\mathcal{P}_k^{\mathrm{poi}}$ denote the sets of benign and poisoned latent codes with source label $k$. We then define BPSS as
\begin{equation}
\mathrm{BPSS}
=
\operatorname{corr}_{\mathrm{S}}
\left(
\operatorname{vec}_{i<j}(\Delta^{\mathrm{ben}}),
\operatorname{vec}_{i<j}(\Delta^{\mathrm{poi}})
\right),
\label{eq:bpss}
\end{equation}
where $\operatorname{corr}_{\mathrm{S}}$ denotes the Spearman rank correlation coefficient, and $\operatorname{vec}_{i<j}$ extracts the strictly upper triangular entries in a common order of class pairs. BPSS thus measures how closely poisoned representations preserve the ordering of benign class-center distances. Because global translation and uniform scaling leave this correlation unchanged, lower BPSS indicates changes in the inherited interclass geometry beyond overall displacement or contraction. Together with the reduced source-class separation observed in latent space visualizations, lower BPSS supports the suppression of residual source information in poisoned representations.

% Queue BPSS separately so the following section can use the right column.
\begin{figure}[!t]
\centering
\includegraphics[width=\linewidth]{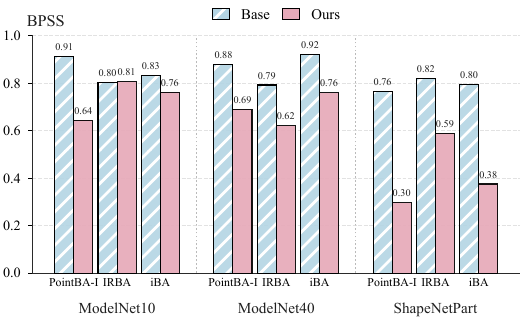}
\caption{BPSS for the analysis of residual source information. We compute BPSS before and after applying AGM across three datasets and three triggers. Ours achieves lower BPSS than Base in most settings, indicating weaker preservation of the benign class-center distance structure in poisoned representations.}
\label{fig:bpss_source_residue}
\end{figure}

Fig.~\ref{fig:bpss_source_residue} shows that our method reduces BPSS in most settings, with more evident drops on ModelNet40 and ShapeNetPart. This decrease indicates that AGM reduces the rank-order agreement between the benign and poisoned class-center distance structures. The main exception is ModelNet10 with IRBA~\cite{gao2024irba}, where BPSS remains nearly unchanged. This observation does not contradict the ASD improvement in Table~\ref{tab:main_results}, since BPSS captures only global distances between class centers rather than all source cues available to the decoder. With only ten coarse categories, ModelNet10 yields 45 class-center pairs, making BPSS more susceptible to stable interclass geometry. Moreover, the WLT module in IRBA preserves coarse source shape through transformations that preserve geometry. AGM can therefore improve attack controllability by suppressing intra-class details, local neighborhoods, or other source cues available to the decoder, even when the global class-center distance structure remains largely unchanged.

%% file: 6_conclusion.tex
\section{Limitations and Future Work}

This study focuses on the poisoning mechanisms of PCAEs and does not quantify the security risks that may arise when poisoned encoders are reused by downstream models. Future work could characterize how representation contamination propagates and persists in self-supervised learning pipelines. It should also assess how the transferability of poisoned representations is affected by the backbone architecture, trigger design, and poisoning rate, and evaluate their effects on downstream 3D representation learning tasks such as classification and point-cloud completion.

\section{Conclusion and Social Impact}

This paper identifies residual source information in poisoned latent representations as a mechanism limiting target reconstruction controllability in PCAEs. The Information Blackhole principle formalizes the objective of retaining target information while suppressing source information, and AGM realizes this objective by matching poisoned representations to an adaptive Gaussian reference. By restricting regularization to the poisoned branch while retaining clean reconstruction supervision, AGM achieves a favorable balance between target reconstruction controllability and benign reconstruction fidelity. In the main experiments, it reduces ASD while keeping CRD largely stable. Latent-space visualizations and BPSS analyses further indicate that AGM attenuates source-class organization in poisoned representations. This latent-space signature could inform audits of potentially compromised 3D reconstruction models and motivate further investigation into the downstream security risks of encoder reuse.